\documentclass[draft]{aipproc}
 \layoutstyle{6x9}

\begin{document}

\title{ Shock-Generated Turbulence  In the Innermost  50 pc  of  the Galaxy Center}
\thanks{Talk presented by Itzhak Goldman}

\author{Itzhak Goldman}{
   address={Department of Exact Sciences, Afeka 
Tel Aviv Academic College of Engineering \\
  Tel Aviv 69107, Israel},   
 altaddress= {School of Physics and Astronomy, Department of Astronomy and Astrophysics, Tel Aviv University 
 Tel Aviv 69978 , Israel} 
 },  
  \author{Marcella Contini}{
address=  {School of Physics and Astronomy, Department of Astronomy and Astrophysics, Tel Aviv University 
 Tel Aviv 69978 , Israel}
 }
\begin{abstract}
  The center of the Milky Way galaxy (MW) is an extreme environment which  contains a massive  black hole surrounded by a very dense star cluster, two other adjacent star clusters, and a giant molecular cloud which would serve as an incubator to a new generation of stars. The  gas and dust in its vicinity are denser by 2-3 orders of magnitude than in other locations in the MW. This is also the case with the magnetic field. The kinematics of the gas is characterized by  apparently random,  supersonic flows. In this paper we provide observational evidence for the existence of a supersonic turbulence, most likely generated by the shock waves. 
Moreover, the mere existence of turbulence and its characteristics are shown to
be instrumental in testing the validity and consistency of theoretical modeling of
the spectra of the gas filaments.
\end{abstract} 

\maketitle 
\section{Introduction}
 
 The center of the Milky Way galaxy (MW) is an extreme environment, very different from the  other locations in  the MW. It contains a dormant $\sim 4\times 10^6 M_{\odot}$ black hole,   three massive clusters of young high-mass stars   and a giant molecular cloud (see reviews in \cite{schdel etal06}). The densities of gas and dust are 2-3 orders of magnitudes higher than the  Galactic average,  the magnetic field  is in the range of  
 $(100\div 1000)\mu G$, while the  Galactic norm is few $\mu G$.  
 The gas exhibits a complex filamentary morphology   as well as a   supersonic velocity field.  
    
   \begin{figure}[h]
$
\begin{array}{cc}
\includegraphics[scale=0.8]{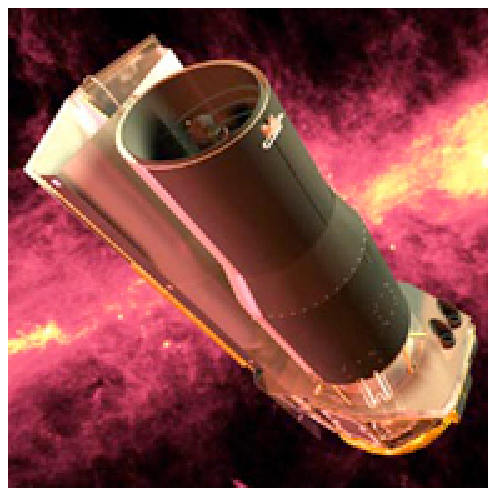}  &
\includegraphics[scale=0.4]{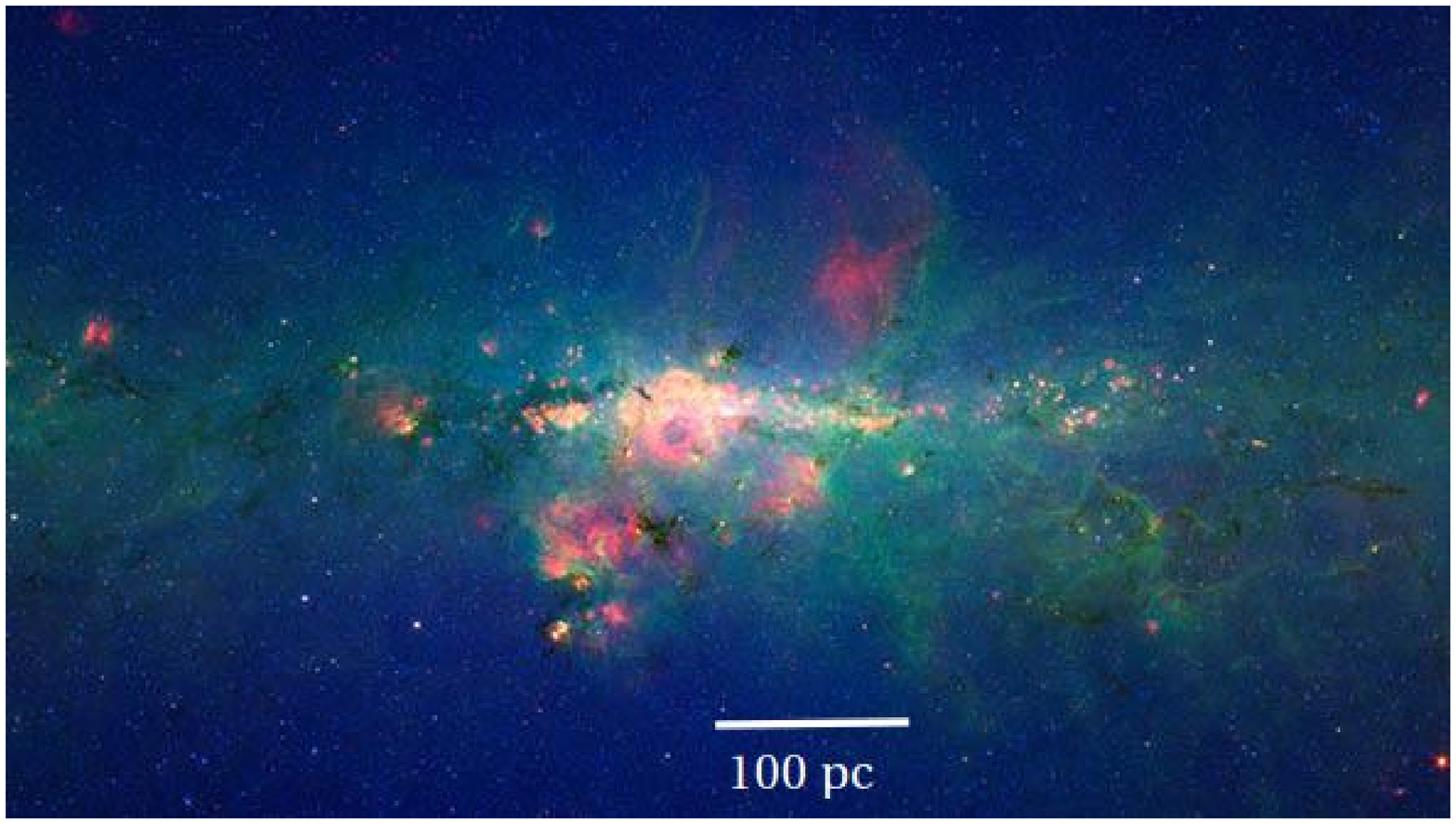} 
\end{array}$
\caption{{\bf left}:  Spitzer Telescope--artist concept. {\bf right}:  Spitzer image of the Galactic center region. {\it blue}: 3.6 $\mu$ m ,{\it green}:   8 $\mu$ m ,{\it red}: 24 $\mu$ m . The brightest white feature at the center of the image is the central star cluster in our galaxy, surrounding the massive black hole. {\it source}:  Spitzer website.} 
\label{GC}
  \end{figure}
  
  The GC has been observed  in  infrared and radio bands because the  high   gas and dust densities combined with a line of sight to the Galactic center (GC)  that passes through the MW disk, strongly absorb  optical  and UV radiation.   Infrared telescopes  have advanced considerably in the past decades;  the Spitzer infrared space telescope is the  most recent and advanced infrared facility to date. The left panel of
   Figure \ref{GC} is an artist concept of the Spitzer telescope while  the right one is  the Spitzer image (in 3 mid-infrared wavelengths) of a $\sim 600 $ pc extent of the MW disk around the GC.

   \begin{figure}[h]
 \centerline{\includegraphics[scale=0.6]{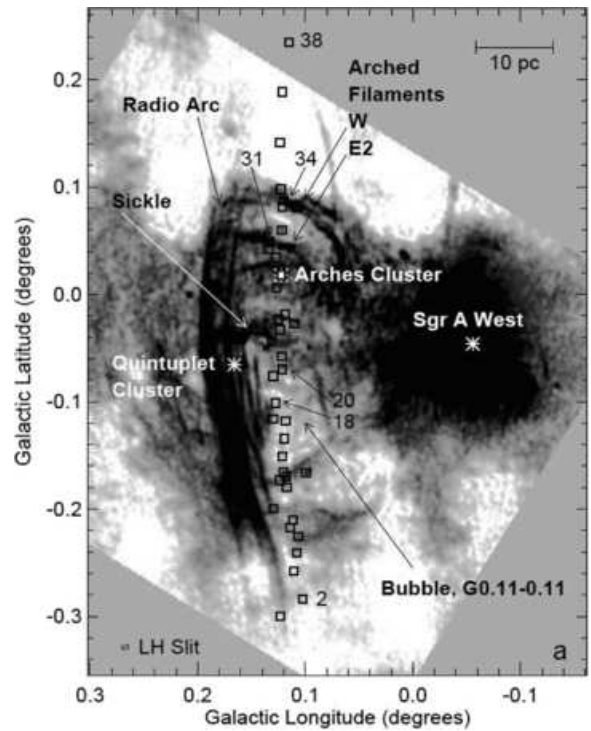} }
\caption{ A 21 cm radio continuum (log-log scale) image  \cite{yusef_morris87}   with 11 arcsec resolution.Except for the Radio Arc and part of the Sgr A region which are non-thermal, all the radio emission is thermal bremsstrahlung. The 38 positions observed  by   with the Spitzer telescope in the mid- infrared \cite{simps07}are indicated by square boxes.(adapted from  Fig. 1 of  \cite{simps07}) }
\label{simps}
   \end{figure}

Zooming in, Figure \ref{simps} which is adapted from  \cite{simps07}, shows  the inner 50 pc of the GC. The black hole Sgr A West is surrounded by the nuclear star cluster (not marked in the figure). At a distance of  about 25 pc  in the Galactic plane   are  the Quintet and the Arches star clusters. Most of the stars are young (age $  \sim  2$ Myr) and massive (10 - 50)~$M_{\odot}$. Visible in the images are the Arches and the Sickle  gas filaments. Visible also is a system of quasi circular arcs whose morphologies suggest  that they were formed by  stellar winds (velocities    $\sim 1000$~km/s) emanating from the   Quintet and the Arches star clusters.

  Observations with the Spitzer Telescope
  by   \cite{simps07}  yielded high quality mid-infrared spectra at 38 positions along a line perpendicular to the MW disk (the positions are marked by the squares  in Figure \ref{simps}). The  observational spectra lead to the conclusion that the gas in the filaments is indeed   excited by ionizing UV photons
  originating in the adjacent  star clusters  \cite{simps07}. However,  photo-ionization by itself cannot account for   the observed line ratios  and an additional excitation by shocks is probably  important. 
Shocks are indeed expected since  
  the line of sight (radial) gas velocities  measured by radio observations (\cite{yusef_etal97, lang_etal05})  span  a range  of 180~km/s.  Moreover, the   filamentary fractured structure of the gas can be naturally explained as the result of turbulence generated by  shocks.

  Following on this proposition, a detailed modeling  
 of the line and continuum spectra allowing for   both photo-ionization and shocks as the sources of excitation has been carried out \cite{contini09} . All the  available lines in each spectrum and the continuum spectral energy distribution were modeled   consistently.  
Indeed, this modeling reproduced correctly the observational spectra  and at the same time was able to derive the  shock velocity, pre-shock density and  magnetic field, and the geometrical depth at each of the 38 positions. More recently \cite{contini_goldman11}, 
these   values of the physical parameters have been used 
  to calculate the expected  spectra in the optical and UV ranges, which   cannot be observed by the   presently available instruments.
  
  \section{Outline}

  The analysis  presented in \cite{contini09}  invoked the existence of shocks as a necessary ingredient
for the  consistent derivation of the spectra observed by   \cite{simps07}.
   Shocks are known to produce turbulence, notably via the Richtmyer-Meshkov instability  (\cite{mikael90},  \cite{graham_zhang}) 
which is due to the shock acceleration, and 
  is similar to the familiar Raleigh-Taylor instability. Shocks can also excite turbulence by creating flows that exhibit 
shear instability or Kelvin Helmholtz instability which in turn  generate turbulence.
Indeed, the clumped and fragmented morphology of the region   strongly  suggests   the existence of an underlying   supersonic shock-generated turbulence.

The existence of turbulence can be probed  directly  by analyzing the velocity spectrum. In addition, the imprint of turbulence may be detected   indirectly  by analyzing the power spectrum of 
"passive scalars"(\cite{lesieurs97}), which are 
strongly coupled to the gas and follow its turbulent motion   (but do  not feed-back on the 
 turbulence). The power spectrum of a passive scalar is
proportional to that of the turbulent velocity.

An example is the   21 cm emissivity which is proportional to the column density of the neutral hydrogen and its fluctuations reflect  the density fluctuations and in turn the velocity fluctuations. In the case of infrared continuum the passive scalar  is the dust density, and in the case of abundances  it is the concentration of the species under consideration. 

The observational  power  spectrum of a passive scalar provides a measure  of its
  hierarchical spatial structure. When this  power spectrum is  a power law, a hydro-turbulence  is naturally suggested as the
mechanism that has shaped the observed spatial    hierarchy. Yet,   by itself, such a spectrum is not enough to prove the existence of a  hydrodynamical turbulence. On the other hand, if the power spectrum of the velocity field reveals a turbulence, the fact that 
 the power spectrum   of a passive scalar is also a power law  with the same  exponent strengthens  the case 
 for the reality  of the velocity turbulence.
  
   An example is the power spectrum of 
21cm emission in the Small Magellanic Cloud (SMC) \cite{stanimir99}. It was shown by \cite{goldman2000} that the power spectrum is consistent 
with that generated by a large scale velocity turbulence. 
Additional support for the existence of a dynamic turbulence was obtained by \cite{goldman07}  by deriving the power spectrum of 
the radial velocities of the giant H$_ I $ super-shells of the SMC and showing that the two power spectra are consistent.
 
In what follows we analyze the radial  velocity data of  \cite{simps07}  and their observed 
mid-IR continuum flux in order to test for the existence of   turbulence. 
Indeed  these two kinds of observational data reveal the existence of a supersonic turbulence.

Following these two analyses,  we examine the  Si/H abundance ratio as function of position  which were {\it  computed} by \cite{contini09} . The idea being that if these computations indeed reproduce   the actual  physical conditions in the region, then
the imprint of the turbulence should be evident in the computed abundance as function of position. 
It turns out to be indeed the case; thus lending 
  credibility to
the computations of \cite{contini09}  as well as to, their extensions  to  the optical and UV  ranges \cite{contini_goldman11}.
Finally, we estimate the effect of turbulence on the magnetic field in the observed positions.

\section{Turbulence: Radial Velocities}

The  3D spectral function of the turbulent velocity, $\Phi(\vec{k})$ is defined in terms of the 2-point
autocorrelation of the turbulent velocity field

$$<\vec{v}(\vec{r'}) \cdot \vec{v}(\vec{r}+\vec{r'})>= \int
\Phi(\vec{k})e^{i \vec{k}\cdot \vec{r}}  d^3k$$
In  the homogeneous and isotropic case it is useful to introduce the
turbulence energy spectrum $E(k)$ and the turbulent velocity spectral function $F(k) = 2 E(k)$ so that

$$\Phi(\vec{k})=\Phi(k)=\frac{  F(k)}{4\pi k^2}\ ;  k=| \vec k|$$
Assuming the ergodic principle, ensemble averages can be replaced by space, surface, or as in the present case, line averages. 

The assumed isotropy implies that $F_r(k)$, - the power spectrum  of the radial turbulent velocity is

$$F_r(k) =\frac{1}{3} F(k)$$

and can be derived from

$$<v_r(x')v_r(x'+x)>= \frac{1}{L}\int_0^L v_r(x')v_r(x'+x)dx = \frac{1}{2\pi}\int_0^L F_r(k)e^{ik x} dk$$
 where $L$ is the length of the line along which the radial velocity has been measured.
The power spectrum thus   is the one dimensional Fourier transform of the 2-point 
autocorrelation function of the radial velocity. Since the data are given at a set of discrete positions we compute the power spectrum by employing a digital Fourier transform. 
Note that at each position  the   mean squared turbulent radial velocity
 
 $$ <v{_r}^2>  = \frac{1}{2\pi}\int_0^L F_r(k)  dk$$
  is contributed by all the spatial scales of the spectrum.

  The observational data are given for 38 positions along an almost straight line
with extension of about 75~pc \cite{simps07}. Not all positions are evenly spaced, thus a numerical
  uncertainty in the small-scales part of the power spectrum is expected.
  
 \begin{figure}[h] 
  \centerline{\includegraphics[scale=1]{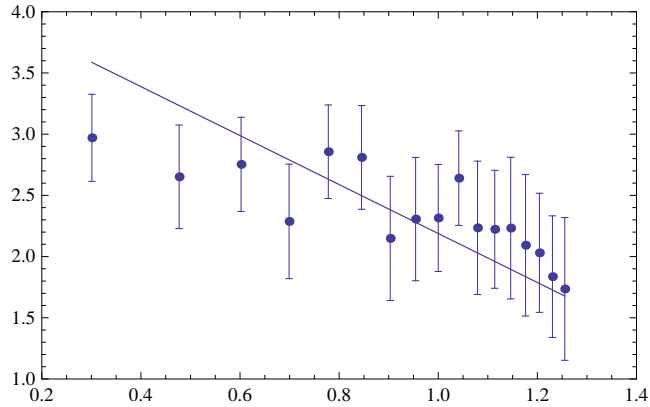}} 
\caption{The power spectrum of the radial velocity residuals, in   units of (km/s)$^2$, as function of the   relative wavenumber k. $k=1$ corresponds to the spatial scale $l_0= 75 pc$.  The line is a power law with index -2.}
\label{vel}
 \end{figure}

Also, the radial velocity observed by \cite{simps07} for the m20 position is substantially larger than the
adjacent velocities and most likely  is a foreground flow. We therefore adopt for this point a value equal to the mean velocity of the other positions.

If the residual velocities  were random {\it uncorrelated} fluctuation  the  power spectrum would have been a "white noise" with the same power on all wavelengths. 
In contrast,
the power spectrum shown in Figure \ref{vel} is a rather steep decreasing function of the wavenumber $k$.
The largest turbulence scale corresponding to the dimensionless wavenumber $k=1$ is the linear extent of the data strip:   $l_0=75$~pc.
The value of the root mean squared turbulent radial velocity at the largest scale is $v_0 = 17.4$~km/s corresponding to a 3D velocity of 30~km/s, well above the   
thermal velocity of $\sim 10$~km/s.

The line in the figure is a $k^{-2}$ power law that is expected for supersonic turbulence (\cite{passot88}, \cite{girimaj95}). It differs from
$k^{-5/3}$ Kolmogorov spectrum which characterizes incompressible turbulence. The steeper slope
is due to the fact that a fraction of the turbulent kinetic energy density  is converted to compression work and is diverted from the energy cascaded to the smaller spatial scales.

 In order to obtain a quantitative estimate of the goodness of the fit we adopted the uncertainty 
of  15~km/s  for each individual measurement \cite{simps07}. Then we obtained synthetic velocity data 
by   generating at each position    a random  velocity using a normal distribution with a mean equal 
to the observed value,  and a standard deviation of  15~km/s. 

We have generated  100 such simulated data sets. For each,   the power spectrum was computed enabling an estimate for the standard deviation of the power spectrum. 

Doing so, a value of 0.72 was obtained for  the reduced $\chi^2$ . Some of the uncertainty in the power spectrum stems from
the uneven spacing of the positions leading to a scatter in the power spectrum at the small spatial scales (large k). The fact that even so the fit is a good one  suggests  that 
the true uncertainty in the observed velocities is {\it smaller} than the assumed value of 15~km/s .

The characteristic timescale of the turbulence is given by $\tau_0\sim l_0/v_0 \sim 4 \times 10^6\ yr$ -- 
  comparable to the age of the young stellar populations. 
This is consistent with the framework adopted  in \citep{simps07} and \cite{contini09} that 
stellar outflows are the main energy  source that shaped  the interstellar medium in this region and together with the stellar UV radiation are also responsible for the observed   mid-infrared  spectra .

The derived  power spectrum  seems to steepen  to a $k^{-3}$ dependence at about $k=7$. If real, it implies that geometrical depth along the radial  line of sight
 is about $\frac{1}{15}$ of the extent on the plane of the sky (\cite{goldman2000}); namely a depth of $\sim 5\ pc$.  
 
\section{Turbulence: Mid-IR Flux}
 
 The observed flux is    an    integral of the emissions along the line of sight. In the optically thin case
it is contributed by all the geometrical depth along the line of sight while in the optically thick case only the foreground layers contribute. Recently,   \cite{mivil07, mivil10}
 have studied fluctuations of the far-IR continuum flux from thermally heated dust  and concluded that their power spectrum is identical to that of the underlying  velocity turbulence. The interpretation of this result was that the dust is strongly coupled to the gas and thus the dust density fluctuations are determined by the gas velocity field which is turbulent. In the cases considered by these authors the turbulence was subsonic and the power spectrum was the Kolmogorov spectrum.
 
Pursuing a similar analysis, we compute the power spectrum of  the  fluctuations of the   mid-IR continuum flux  in the   range $(13.5-14.3)\mu m$, as function of position.  Mathematically, the procedure is the same employed in the previous section with regard to the radial velocity.

\begin{figure}[h]  
\centerline{\includegraphics[scale=1]{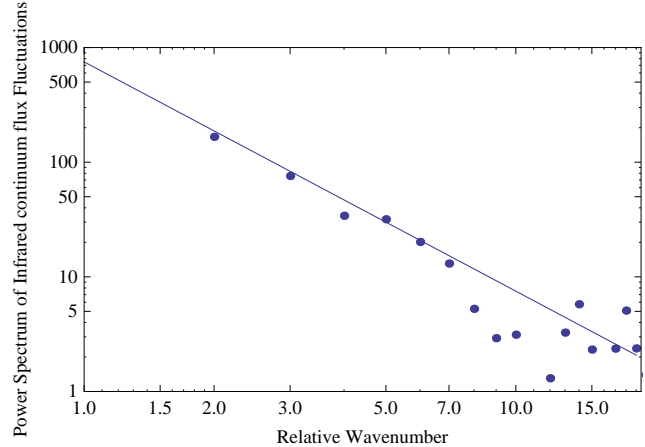}}
\caption{The power spectrum of the mid-IR continuum at $(13.5-14.3)\mu m$, in   units of Jy$^2$, as function of the   relative wavenumber k. $k=1$ corresponds to the spatial scale $l_0= 75 pc$.  The line is a power law with index -2.}
\label{ir}
 \end{figure}

The resulting power spectrum is displayed in Figure \ref{ir}. It exhibits a clear $ k^{-2}$ behavior  for the larger scales and   an indication of a possible steepening at about $ k=7$ to a $k^{-3}$ dependence; in effect the latter  is more obvious here than in the radial velocity spectrum. 
 
The uncertainties reported by  \cite{simps07}  are 
 quite small (of the order of $1\%$ ). As a result the uncertainties in the power spectrum 
 are too small to be indicated in the figure. The scatter in the power spectrum is dominated by the  uneven spacing of the positions \cite{simps07}. The latter can be estimated 
  by generating synthetic data with randomly spacings 
(much in the way as with  the velocity uncertainties in the previous section),
However this seems to be outside the scope of the present paper. 
Therefore, we do not derive here a quantitative estimate for the goodness of fit. 
Rather we would like to draw attention to the fact that the deviations from the power law are  mostly at the small  spatial 
scales, as expected from an uneven spacing, and not on the large  scales.

 The important conclusion from this power spectrum  is that it is consistent with the interpretation that the dust density fluctuations (responsible for the flux fluctuations) were driven by the    supersonic velocity turbulence addressed in the previous section. 
 
\section{Turbulence: Si/H  Abundance Ratio}
 In \cite{contini09}  various abundance ratios were computed as function of position.
If these computations are a fair representation of the physics in the region, it is  expected that the computed abundances will reflect the turbulence revealed in the observational radial velocity and mid-IR flux.

 If the heavy elements, whose abundance is considered, are 
strongly coupled to the gas and follow the gas motions then they can be regarded  as a  turbulent "passive scalars" (\cite{lesieurs97}), much in the same way as the dust density responsible for the mid-IR emission.   As in the latter case, the power spectrum   should be 
proportional to the turbulent velocity power spectrum.  
We   computed the power spectrum 
 of  the  fluctuations of the Si/H ratio  at the 38 positions mentioned in the previous sections. This specific abundance  was selected   because it does not show drastic variations with position and thus  (unlike as with e.g. Fe/H)  the fluctuations are not affected by  trapping into dust grains and eventual sputtering. 

\begin{figure}[h]
 \centerline{\includegraphics[scale=1]{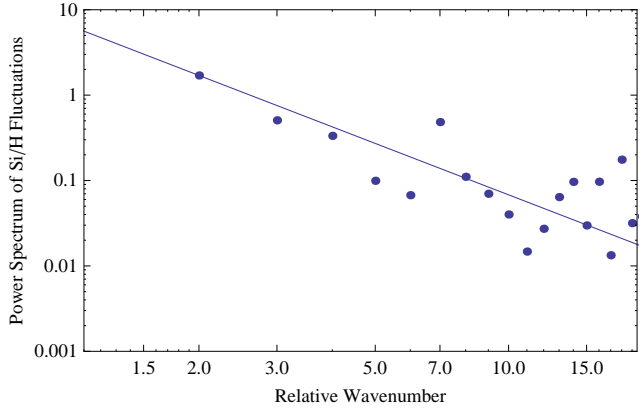}}
\caption{ The power spectrum of Si/H residuals, in  units of $10^{-10}$,  as function of the relative wavenumber k. $k=1$ corresponds to the spatial scale $l_0= 75 pc$.  The line is a power law with index -2.}
\label{si}
 \end{figure}

 The resulting power spectrum is shown in Figure \ref{si}. The uncertainties in the power spectrum resulting from the uncertainties of the computed abundance are too small to be indicated in the figure.  The scatter seen in the small scales (large wavenumbers) probably reflects the uneven spacing of the observational positions.
The power spectrum is consistent with the $k^{-2}$ spectrum of the radial velocity turbulence and that of  the mid-IR flux. The consistency of the power spectrum with what is expected from an underlying supersonic hydrodynamic turbulence strengthens 
the case for the existence of the latter. 
More importantly, this lends credibility to the compuational model employed in \cite{contini09}.     
Here, too there is an indication for  a steepening of the spectrum at $ k \sim 7$ to a $k^{-3}$ dependence.\\

\section{Discussion}

 The clumped and fragmented morphology of the GC region   strongly  suggests that it has been shaped
 by a supersonic turbulence. The latter has been  generated by  shocks driven by the high velocity stellar winds from  the young massive stars  located in the adjacent star clusters. 
 Shocks are known to produce turbulence directly, via the Richtmyer-Meshkov instability  
  due   to the shock acceleration, and also indirectly by giving rise to 
shear  and Kelvin Helmholtz instabilities which in turn excite turbulence.
 
 The existence of such a turbulence was revealed in  the power spectra of the 
observational radial velocities and mid-IR continuum flux. The power spectra exhibit a $k^{-2}$ behavior typical to supersonic turbulence, and is steeper 
than the Kolmogorov spectrum that corresponds  to incompressible turbulence.
A steepening of the power spectra for relative wave numbers exceeding 7, suggests that the line of 
sight depth of the turbulent region is about 5pc.
The turbulence radial root mean squared velocity is about 17~km/s and its 3D value is about 30~km/s indicating a turbulence Mach number of about 2-3.
 The associated turbulence timescale is about 4 Myr -- comparable to the age   of the young 
stellar populations which are  ultimate generators of the shocks and the ensuing turbulence. 
 Since the modeling of the observed mid infrared spectra \cite{contini09}  invokes the existence of shocks-- our finding 
  lends credibility to the latter assumption. The  models  also predict  the   abundances as function of position \cite{contini09}. Analysis of the Si/H abundance  field,  reveals the same underlying turbulence as do the observational velocities and the mid-infrared flux. This implies that the modeling of \cite{contini09}  does indeed reproduces the true physical
  parameters of the gas.

 The turbulence can amplify the initial magnetic fields \cite{contini09} up to equipartition value  if the available time is much larger than the turbulence timescale. The amplification 
proceeds via the dynamo mechanism involving the winding up of magnetic field flux lines by the turbulence eddies.  
 
  For the 3D   turbulent velocity of  30~km/s , the equipartition value for  the largest scale is  
 $B_{turb,eq}=140 \mu G \  (n/ 100cm^{-3})^{1/2}$ 
with $n$ denoting the gas number density.  
 
Fields of this strength and even an order of magnitude higher were reported for the Galactic 
center region (see review by \cite{Val} ).  The  larger fields could have evolved from the turbulently 
amplified field by an inverse cascade process (\cite{mininni07}) from small to large spatial scales, or directly by shock compression of the
magnetic flux lines embedded in the
ionized ISM \cite{medvedev07}.

\begin{theacknowledgments}
Itzhak Goldman thanks Michael Mond and Pierre-Louis Sulem for the invitation to   the interesting and well organized WISAP-2011 conference in Eilat. He  also wishes to thank
the Afeka College Research Authority for
financial support.  
\end{theacknowledgments}

\bibliographystyle{aipproc}

\end{document}